\documentstyle[hip-artc,epsfig]{article}  
\volnumber{9}  \edyear{1999}  \frompage{000} \topage{000}                
\recrevdate{1 January 1999}                                              
\def\rightmark{(Short) Title of the Article} 
\def\leftmark{Author One et al.}
\title{Fission Characteristics of Heavy Nuclei:\\ Statics and Dynamics} 
\authors{
{\twerm Birger B. Back}
\\[2.812mm]
{\normalsize
Argonne National Laboratory\\ 
Argonne, IL 60439, U.S.A.\\[0.2ex] }
}

\abstract{This paper presents a selective historical perspective
of fission research over the last thirty-five years while Ray Nix has made
central contributions to the field. The emphasis is placed
on early studies of the shell stabilized secondary minimum in the static 
fission barrier and on the 
dynamic properties of fission of hot nuclei, which have 
recently been the focus of intense study.}
 
\begin{document}
 
\maketitle

\section{Introduction}

Ray Nix's contributions to nuclear fission research span his whole
career and a wide range of topics. As the title indicates, these
include both the static properties of fissioning systems, such as the
potential energy surface as a function of various shape parameters, as well
as the dynamic aspects of fission, which traditionally has been studied in
terms of the kinetic energy release. Recently it has, however, become
possible to measure the time scale of fission and learn about the effects of
friction or viscosity of nuclei undergoing large shape changes. 

In this talk I would like to review just a few of these topics which are
selected only on the basis that they overlapped with my own work.

\section{The double-humped fission barrier}

In the late sixties, there was much excitement in the field of nuclear
fission research when it was realized that several puzzling experimental
results found a consistent explanation \cite{Strutinski_Bjornholm68} in
terms of a secondary minimum in the fission barrier. This is
caused by the shell
stabilization at large deformation with an axis ratio of 2:1 of axially
symmetric prolate deformation. The experimental evidence included the
observation of anomalously short-lived fission activities in the bombardment
of $^{238}$U with beams of $^{22}$Ne and $^{16}$O \cite{Polikanov62}, the
observation of an enhanced fission decay width at intervals of about 650 eV
in slow neutron induced fission of $^{240}$Pu \cite{Migneco68} as well as
gross resonances in the $^{230}$Th(n,f) \cite{Vorotnikov67}, 
$^{233,235}$U(d,pf) and $^{239,241}$Pu(d,pf) \cite{Pedersen69} reactions.
Realizing that the the fission barrier in actinide nuclei typically is
double-humped, these observations were naturally explained as manifestations
of spontaneous fission of nuclei trapped in the super-deformed minimum and
the enhancements in the fission decay width afforded by the coupling to
compound states and `$\beta$'-vibrational resonances in the second well,
respectively. In fact, the vibrational
resonance in the fission excitation function of $^{239}$Pu(d,pf) is clearly 
observed in a decade earlier experiment by Northrop, Stokes, and Boyer 
\cite{Northrup}, but it was mis-interpreted as a plateau created by a wide gap 
between the lowest fission channels.  

\begin{figure}[htb]
\center{
\vspace*{0.5cm}
\epsfig{file=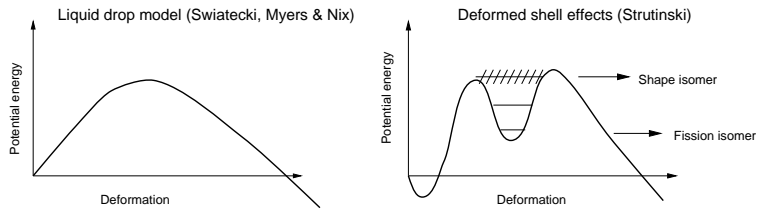,width=10cm}
\caption[]{
The liquid drop fission barrier \cite{Swiatecki} (top panel) is modified by
the deformation dependence if the single particle shell structure
\cite{Strutinski}. The double-humped potential gives rise
to short-lived fission isomers and shape resonances as discussed in the text.
}}
\label{fig1}
\end{figure}

With the double-humped barrier well established in the actinide region it
quickly became clear that further detailed analysis of data required fast
and reliable algorithms for calculating the quantum mechanical tunneling
through this complex barrier. Different
parameterizations of the double-humped shape were studied and both analytical
\cite{Cramer_Nix}
and numerical \cite{Bondorf} solutions to the quantum-mechanical tunneling 
problem were found. Cramer and Nix proposed \cite{Cramer_Nix} a
parameterization in terms of three smoothly joined parabolas, which has a
sufficient number of parameters to describe the essential characteristics of
the deformation path and was associated with a relatively simple analytical
solution to the tunneling problem. 

These calculations all showed the peaks
in the tunneling probability associated with vibrational states in the
second well, but it soon became clear that the large observed width of the 
resonances could not be reproduced \cite{Back69}.
Bondorf had pointed out, however,  that this is a manifestation of the
coupling between the purely vibrational states and the
much more abundant compound states in the second well, and that this effect
can be included by introducing an absorptive imaginary potential 
\cite{Bondorf}. 

Having
arrived at Los Alamos in the spring of 1971 to work with Chip Britt,
I wanted to analyze our fission probability data with this resonance 
structure using Ray's method but needed to include the imaginary potential
to correctly reproduce the observed resonance widths. 

\begin{figure}[htb]
\center{
\epsfig{file=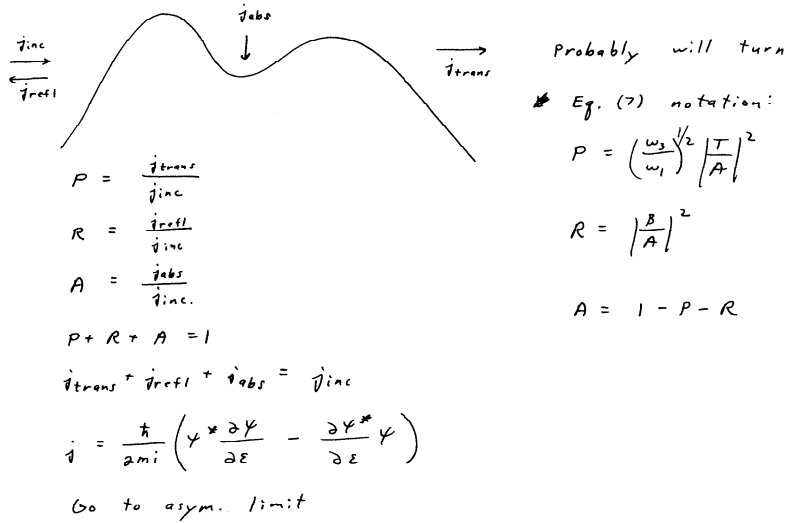,height=7cm}
\caption[]{
Notes on the calculation of the transmission through the double-humped
barrier in the presence an absorptive negative imaginary potential in the
second well \cite{Nix_notes} 
}}

\label{fig2}
\end{figure}

Attempting to consult with Ray on this topic I went to his office.
Initially I thought that it was vacant - the desktop was clean,
without the normal stacks of books and papers, but upon closer inspection I
discovered one open book on the desk and neatly arranged ring binders and
books on the shelves. Later I learned that this was quite normal. I also 
managed to talk to Ray about the problem of
absorption in  the second well and soon received the note shown in Fig. 1,
which indicates how the solution can be obtained. 

\begin{figure}[htb]
\center{
\epsfig{file=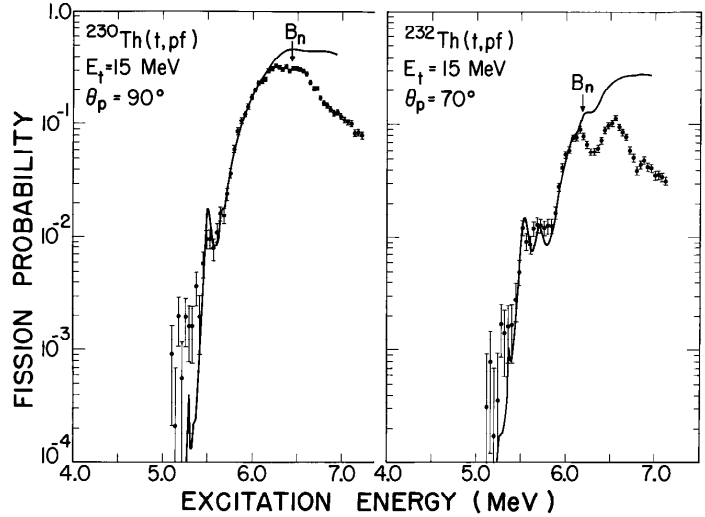,height=7cm}
\caption[]{Fission probability for the $^{230,232}$Th(t,pf)-reactions
\cite{Back72}.
}}
\label{fig3}
\end{figure}

An example of fission probability data is
shown in Fig. 3 for the $^{230,232}$Th(t,pf) reactions \cite{Back72}. 
Strong vibrational resonances are clearly
visible - in particular in the the $^{232}$Th(t,pf) reaction - and the data
are quite well reproduced within the model calculations. A large number of
nuclei in the actinide region were studied in this manner and from the
analysis it was possible to determine
the heights of both the inner and outer barriers, when combined with
an analysis of fission isomer excitation functions and half-lives \cite{Britt73}.
The systematics of the fission barriers heights for even-even isotopes of
Th, U, Pu and Cm obtained in this manner is shown in Fig. 4 . We observe that the
inner barrier height, $E_A$ (top panels) is rather constant over this range
of nuclei whereas the outer barrier height $E_B$ (bottom panels) falls off 
with atomic number of the fissioning system. 

\begin{figure}[htb]
\center{
\epsfig{file=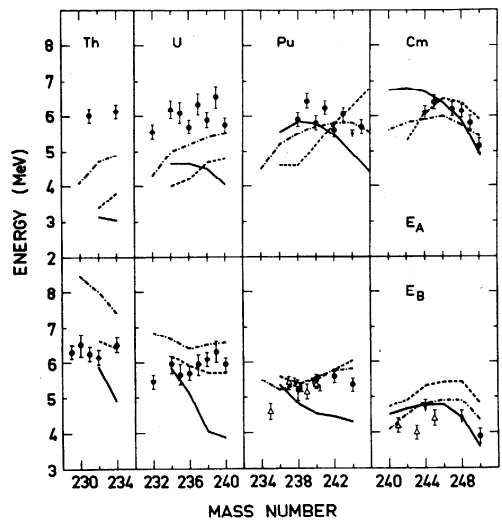,height=7cm}
\caption[]{Experimental fission barrier heights for even-even Th, U, Pu,
and Cm isotopes \cite{Back72} (solid circles) \cite{Britt73} (open triangles) 
are compared to Strutinski-type calculations using a folded Yukawa (solid
curves) \cite{Moller_Nix}, a modified harmonic oscillator (dashed-dotted curves)
\cite{Moller_Nix}, and a Woods-Saxon potential (dotted curves) \cite{Pauli}.}
}
\label{fig4}
\end{figure}

The solid and dash-dotted lines
are calculations by M\"{o}ller and Nix \cite{Moller_Nix} using a folded Yukawa
and a modified harmonic oscillator potential, respectively, whereas the
dotted curve was obtained with a Woods-Saxon potential by Pauli and
Ledergerber \cite{Pauli}. The most striking discrepancy between the data and
calculations is seen for the inner barrier in Th. It has later been verified
that this discrepancy is only apparent because the barrier heights obtained from
the fission probability data are in fact those of a double peaked outer
barrier; the inner barrier has not yet been measured and may well be as low
as predicted.

\section{Fission dynamics - time-scales}

For many years, the empirical information on fission dynamics
was obtained almost exclusively from studies of
the total kinetic energy ($TKE$) release in fission. This quantity
was shown to follow a fairly simple systematics \cite{Viola},
with an approximately linear
dependence on the parameter $Z^2/A^{1/3}$, which scales with the Coulomb
repulsion between the nascent fragments at scission. A linear scaling
between $TKE$ and $Z^2/A^{1/3}$ is obtained if one assumes that 1) the system
is essentially stationary at scission and 2) that the overall shape (charge
distribution) is independent of the size of the system. However, as shown by
Davies, Sierk, and Nix \cite{Nix_TKE} these conditions are not necessarily fulfilled in
realistic dynamical calculations 
within the liquid drop model incorporating a viscosity term. They found,
however, that for a viscosity of $\eta$ = 0.015 TP (tera poise) (water has a
viscosity of about $\eta \sim$ 0.01 P at room temperature) could account 
well for the total kinetic energies over the range of available data. 

\subsection{Pre-scission neutron and $\gamma$-ray emission}

\begin{figure}[htb]
\center{
\epsfig{file=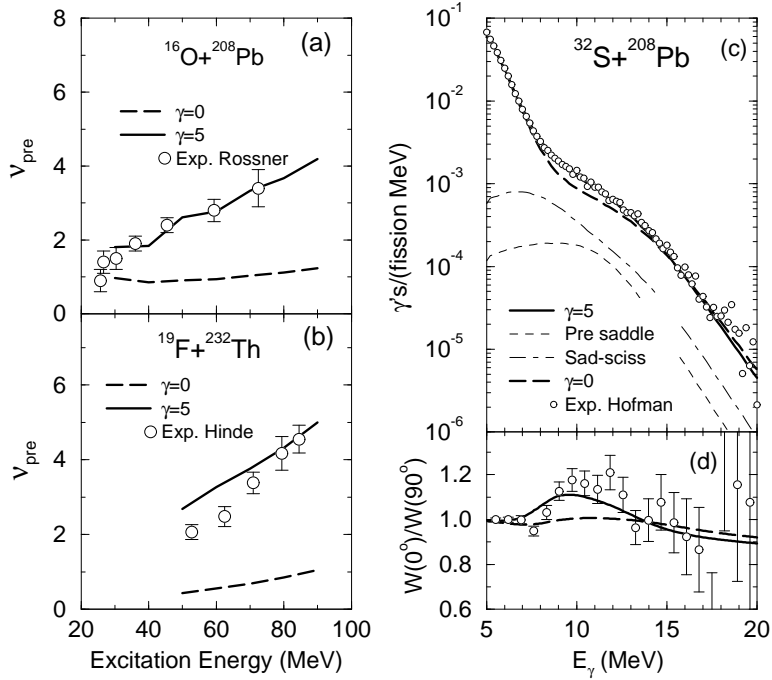,height=9cm}
\caption[]{Pre-scission neutron multiplicities (open circles) 
for the (a)  $^{16}$O+$^{208}$Pb \cite{Rossner} and (b) 
$^{19}$F+$^{232}$Th \cite{Hinde} reactions are compared to statistical model
calculations with (solid line) and without (dashed line) dissipation.
The $\gamma$-ray spectrum (c) and anisotropy (d) in coincidence with fission
fragments \cite{Hofman} are compared to statistical model calculations.}
}
\label{fig5}
\end{figure}

In a set of pioneering experiments by Gavron {\it et al.} \cite{Gavron},
and later by Hinde {\it et al.} \cite{Hinde},
it was found that the multiplicity of neutrons emitted prior to scission
in hot nuclei was substantially larger than expected for the time-scales
for saddle-to-scission motion of $t_{ss}$ = 3.5$\times$10$^{-21}$ seconds
predicted by these dynamical calculation. This is illustrated in
Fig. 5a and 5b, where the observed (open circles) pre-scission neutron
multiplicity, $\nu_{pre}$, exceeds the prediction from a purely
statistical decay model (dashed curves); the short saddle-to-scission time of
 $t_{ss}$ = 3.5$\times$10$^{-21}$ seconds would allow only for the
additional emission of a fraction of a neutron during the descent from
saddle to scission. Also the pre-scission emission of $\gamma$-rays in the
Giant Dipole Resonance region (low energy component) of $E_{\gamma}$ = 9-13
MeV is seen to exceed the expectation from the statistical model (Fig. 5c). 
This also
manifests itself in an enhanced angular anisotropy of $\gamma$-ray emission
relative to the normal of the reaction plane containing the two fission
fragments, as illustrated in Fig 5d.

The full drawn curves in Fig. 5 all represent calculations which include the
effects of dissipation in the fission process. Instead of counting up
transition states at the fission saddle point, which leads to the standard
Bohr-Wheeler \cite{Bohr-Wheeler} expression for the fission decay width, we
have here used an expression based on a diffusion picture for the fission
originally proposed by H. A. Kramers \cite{Kramers}. The fission decay width
is thus assumed to be given by

\begin{equation}
\Gamma_f(t) = \Gamma_f^{BW}\{1-\exp(-t/\tau_D)\}\{\sqrt{1+\gamma^2}-\gamma\},
\end{equation}

\noindent where the factor $\{1-\exp(-t/\tau_D)\}$
takes into account the fact that the buildup in fission
flux over the saddle point occurs over a period of $\tau_D$, whereas the
factor $\{\sqrt{1+\gamma^2}-\gamma\}$, the so-called Kramers factor,
represents the reduction in the asymptotic fission width caused by
dissipation. The normalized linear friction
coefficient $\gamma = \beta/2\omega_0$ is given in terms of the reduced
dissipation coefficient, $\beta$, and the characteristic frequency
$\omega_0 \sim 1 \times 10^{21} s^{-1}$ associated with
the curvature of the saddle point. In addition to the delay and reduction of
the fission flux across the saddle point, the dissipation also lengthens the
time, $\tau_{ss}$, for the descent from saddle to scission according  to the
relation \cite{Hoffman-Nix}
\begin{equation}
\tau_{ss} = \tau_{ss}^0\{\sqrt{1+\gamma^2}+\gamma\}
\end{equation}
where $\tau_{ss}^0$ is the saddle-to-scission time without dissipation in
the system.
For the calculations shown in Fig. 5 as solid  curves, a value of $\gamma$ =
5 was used, which is seen to account reasonably well for the observed
pre-scission of neutrons and $\gamma$-rays. However, the data for the
$^{19}$F+$^{232}$Th reaction gives a hint that the dissipation strength is
somewhat weaker at the lower excitation energies.

\subsection{Cross sections}

It is also expected that the fission dissipation will have a substantial
effect on the competition between fission and particle evaporation in the
decay cascade of hot nuclei. This will manifest itself in the observed
cross section for fission or evaporation residue formation. In Fig. 6 this
effect is illustrated for three different experiments, namely the
evaporation residue cross section for $^{32}$S+$^{184}$W, the survival
probability of Th-like nuclei formed in  deep-inelastic collisions between
400 MeV $^{40}$Ar and $^{232}$Th, and the fission cross section for the
$^3$He+$^{208}$Pb reaction.  

\begin{figure}[htb]
\center{
\epsfig{file=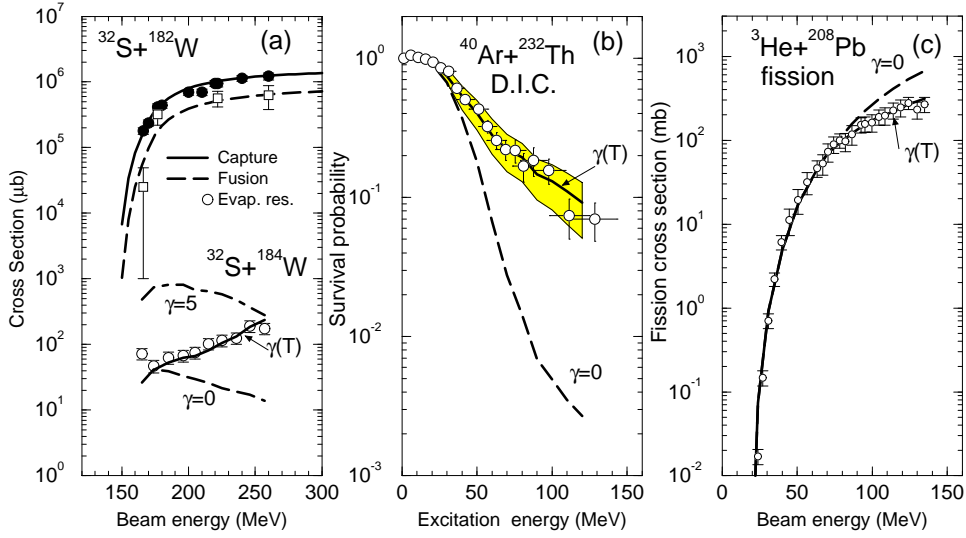,width=\textwidth}
\caption[]{Panel a: Evaporation residue cross sections for the
$^{32}$S+$^{184}$W reaction (open circles) \cite{Back99} is compared with
statistical model calculations (see text).
The fission (solid circles) and complete fusion (open squares) cross sections
for $^{32}$S+$^{182}$W \cite{Back96} are also shown. Panel b: The survival
probability (open circles) of Th-like residues from the
$^{40}$Ar+$^{232}$Th deep inelastic reaction at 400 MeV beam energy
\cite{Hofman99}. Panel c: Fission cross section for $^{3}$He+$^{208}$Pb
\cite{Rubehn}.}
}
\label{fig6}
\end{figure}

We observe in Fig. 6a that the cross section for
evaporation residues in $^{32}$S+$^{184}$W collisions \cite{Back99}
increase
with beam energy contrary to the purely statistical model prediction of a
decreasing excitation function illustrated by the long-dashed curve labelled
$\gamma$=0. When a fixed dissipation strength of $\gamma$=5 is introduced, 
the predicted evaporation residue cross section is increased by
a factor of ten (dashed-dotted curve labelled $\gamma$=5 in Fig. 6a).
However, none of these calculations reproduce the increase in cross section
with excitation energy, and it appears that a temperature
(excitation energy) dependent dissipation is required by the data. This observation
has been made earlier by Hofman {\it et al.} \cite{Hofman95} 
on the basis of pre-scission
$\gamma$-ray measurements at different beam energies. The solid curve
labelled $\gamma(T)$ incorporates a temperature dependent dissipation
strength of the form displayed in Fig. 7 (open triangles) and is seen to
give a good account of the measurements.

A similar conclusion is drawn from an analysis of the survival probability
of Th-like recoils from deep inelastic scattering shown in Fig. 6b. Here the
probability for survival of the recoiling target-like reaction partner
associated with detected Ar-isotopes is plotted as a function of excitation
energy. The correspondence between the measured reaction Q-value and the
excitation energy of the target-recoil is not straight forward and is
discussed in more detail in Ref. \cite{Hofman99}. We observe that the
survival probability does not decrease as precipitously with excitation
energy as predicted by the standard statistical model represented by the
long-dashed curve labelled $\gamma$=0. On the contrary,
a dissipation strength that increases with
temperature (solid curve, $\gamma(T)$), as shown by the open squares in Fig.
7, is required in order to account for the observed survival probability
(solid curve, $\gamma(T)$, Fig. 6b).

Alternatively, it is equally instructive to analyze the fission cross
section in a lighter system such as $^3$He+$^{208}$Pb \cite{Rubehn}, 
where fission
is a relatively weak decay branch. We observe in Fig. 6c, that the measured
fission cross section (open circles)
does not increase as rapidly with beam energy as
expected from the statistical model (long-dashed curve). 
Again, a dissipation strength which increases with excitation energy as shown 
by open circles in 
Fig. 7, is required to reproduce the measured cross section 
(solid curve, $\gamma(T)$, Fig. 6c).

\begin{figure}[htb]
\center{
\epsfig{file=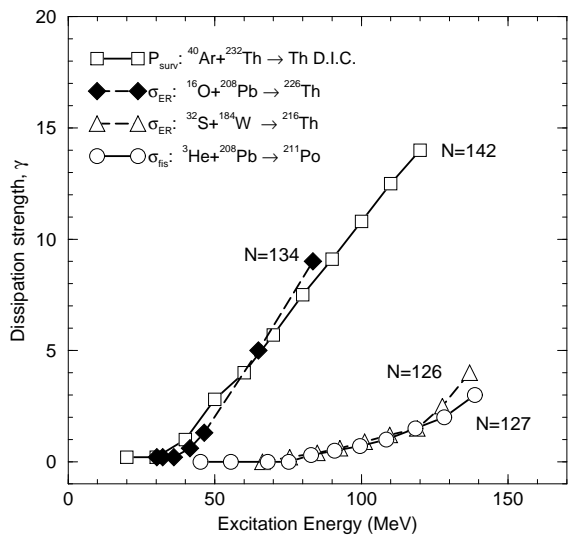,height=7cm}
\caption[]{Extracted dissipation strength derived from different
experimental data}
}
\label{fig7}
\end{figure}

The excitation energy dependence of the dissipation strength, $\gamma$,
obtained from the above analysis is
summarized in Fig. 7. Although it is clear that dissipation strengths
increasing with temperature appears to be a general result from this
analysis, it is interesting to note that the rate of increase varies strongly
from system to system. The four systems studied here appear to divide into
two pairs. The two  Th-systems with masses of A=226 and A$\sim$232
corresponding to neutron numbers of N=134, and $\sim$142 exhibit a rapid increase
in the dissipation strength above an excitation energy of $\sim$40 MeV,
whereas the $^{216}$Th and $^{211}$Po systems exhibit little or no
dissipation up to an excitation energy of about 80 MeV followed by a gentle
increase. The two latter systems have closed
neutron shells  (N=126 and N=127) hinting at a dependence on the shell
structure of the fissioning system. Thus, it appears that a closed nuclear
shell structure suppresses the fission dissipation up to excitation energies
of E$_{exc}\sim$80-100 MeV, where the shell structure itself is dissolved.
Admittedly, these conclusions are based in a very sparse data sample, and it
would be very interesting to perform a more systematic study of these
effects which spans a wider region with varying shell structure.

\section{Conclusions}

The continued interest in the fission process arises from the intricate
relationship between the dynamics of large scale shape evolutions of a
nuclear system and the underlying static properties.  Over the last thirty-five
years the study of this relationship has progressed
very rapidly. In the early part of the period the research was concentrated
on the static nuclear properties, such as the potential energy surface as a
function of deformation. This focus was brought about by the
the discovery of the complexity of the
fission barrier and the many consequences this entailed for the fission 
properties of
relatively cold nuclei. In the latter part of this period the studies
shifted to fission of highly excited and/or rapidly rotating nuclei formed in
reactions with heavy ion beams. In this talk I have focussed on a few of the
many important contributions to this study that Ray Nix has made during his
distinguished career in nuclear physics.

\section*{Acknowledgement}
This work was carried out under the auspecies of the U. S. Department of
Energy under contract No. W-31-109-Eng38.

\vfill\eject

\begin{thebibliography}{99}  

\bibitem{Strutinski_Bjornholm68}
        S. Bj\/{o}rnholm and V. M. Strutinski, 
	Nucl. Phys. {\bf A136} (1969) 1.

\bibitem{Polikanov62} 
        S. M. Polikanov, V. A. Druin, V. A. Karnaukov,
        V. L. Mikheev, A. A. Pleve, N. K. Skobelev, 
	V. G. Subbotin, G. M. Ter-Akopyan, and V. A. Fomichev,
	J. Exptl. Theoret. Phys. (U.S.S.R.) {\bf 42} (1962) 1464; 
	[Transl: Sov. Phys JETP, {\bf 15} (1962) 1016]

\bibitem{Migneco68}
        E. Migneco and J. P. Theobald,
	Nucl. Phys. {\bf A112} (1968) 603.


\bibitem{Vorotnikov67}
        P. E. Vorotnikov, S. M. Dubrovina, V. A. Shigin,and G. A. Otroschenko,
	Sov. J. Nucl. Phys. {\bf 5} (1967) 210.

\bibitem{Pedersen69}
        J. Pederson and B. P. Kuzminov, 
	Phys. Lett. {\bf 29B} (1969) 176;
	B. B. Back, J. P. Bondorf, G. A. Otroschenko, J. Pedersen, and
	B. Rasmussen, Nucl. Phys. {\bf A165} (1971) 449.
	
\bibitem{Northrup}
        J. A. Northrop, R. H. Stokes, and K. Boyer,
	Phys. Rev. {\bf 115} (1959) 1277.
	
\bibitem{Swiatecki}
        J. R. Nix and W. Swiatecki,
	Nucl. Phys. {\bf 71} (1965) 1;
        W. D. Myers and W. Swiatecki,
	Nucl. Phys. {\bf 81} (1966) 1;
	J. R. Nix, Nucl. Phys. {\bf A130} (1969) 241.
	
\bibitem{Strutinski}
        V. M. Strutinski, Nucl. Phys. {\bf A95} (1967) 420;
	V. M. Strutinski, Nucl. Phys. {\bf A122} (1968) 1.

\bibitem{Cramer_Nix}
        J. D. Cramer and J. R. Nix,        
	Phys. Rev. {\bf C2} (1970) 1048;
	C. Y. Wong and J. Bang,
	Phys. Lett. {\bf 29B} (1969) 143.
		
\bibitem{Bondorf}
        J. P. Bondorf, Phys. Lett. {\bf 31B} (1970) 1.

\bibitem{Back69}
        B. B. Back, J. P. Bondorf, G. A. Otroschenko, J. Pedersen, and
	B. Rasmussen, Proc. of Physics and Chemistry of Fission, 
	(IAEA, Vienna, 1969) p. 351;

\bibitem{Nix_notes}
        Notes by Ray Nix given to me, June 1971.

\bibitem{Back72}
        B. B. Back, H. C. Britt, J. D. Garrett, and Ole Hansen,
	Phys. Rev. Lett. {\bf 28} (1972) 707. 

\bibitem{Britt73}
        H. C. Britt, M. Bolsterli, J. R. Nix, and J. L. Norton,
	Phys. Rev. {\bf C7} (1973) 801.

\bibitem{Moller_Nix}
        P. M\"{o}ller and J. R. Nix,
	in Proceedings of the Third International Conference on the Fission 
	and Chemistry of Fission, Rochester, 1973, (IAEA, Vienna, 1974) p. 103.

\bibitem{Pauli}
        H. C. Pauli and T. Ledergerber,
	Nucl. Phys. {\bf A175} (1971) 545.
	
\bibitem{Viola}
        V. E. Viola, Jr. Nucl. Data, {\bf1} (1966) 391. 

\bibitem{Nix_TKE}
        K. T. R. Davies, A. Sierk, and J. R. Nix,
	Phys. Rev. {\bf C13} (1976) 2385.

\bibitem{Gavron}
        A. Gavron, J. Beene, B. Cheynis, R. L. Ferguson, F. E. Obershain,
	F. Plasil, G. R. Young, G. A. Petitt, M. J\"{a}\"{a}skel\"{a}inen, 
	D. G. Sarantites, and C. F. Maguire,
	Phys. Rev. Lett. {\bf 47} (1981) 1255 [erratum: {\bf 48} (1982) 835]

\bibitem{Rossner}
        H. Rossner, D. J. Hinde, J. R. Leigh, J. P. Lestone, J. O. Newton,
	J. X. Wei, and S. Elfstr\"{o}m,
	Phys. Rev. {\bf C45} (1992) 719.
	
\bibitem{Hinde}
        D. J. Hinde, R. J. Charity, G. S. Foote, J. R. Leigh, J. O. Newton,
	S. Ogaza, and A. Chattejee,
	Hucl. Phys. {\bf A452} (1986) 550.

\bibitem{Hofman}
        D. J. Hofman, B. B. Back, I. Di\'{o}szegi, C. P. Montoya,
	S. Schadmand, R. Varma, and P. Paul,
	Phys. Rev. Lett. {\bf 72} (1994) 470.

\bibitem{Bohr-Wheeler}
        N. Bohr and J. A. Wheeler,
	Phys. Rev. {\bf 56} (1939) 429.

\bibitem{Kramers}
        H. A. Kramers,
	Physiza {\bf 7} (1940) 284.

\bibitem{Hoffman-Nix}
        H. Hoffman and J. R. Nix,
	Phys. Lett {\bf 122B} (1983) 117.

\bibitem{Back96}
        B. B. Back, P. B. Fernandez, B. G. Glagola, D. Henderson, S. Kaufman,
	J. G. Keller, S. J. Sanders, F. Videbaek, T. F. Tsang, and B. D. Wilkins,
	Phys. Rev. {\bf C53} (1996) 1734.

\bibitem{Back99}
        B. B. Back, D. J. Blumenthal, C. N. Davids, D. J. Henderson, 
	R. Hermann, D. J. Hofman, C. L. Jiang, H. T. Penttil\"{a}, and
	A. H. Wuosmaa,
	To be published in Phys. Rev. {\bf C} (1999)
	
\bibitem{Hofman99}
        D. J. Hofman, B. B. Back, D. J. Henderson, V. Nanal, and 
	A. H. Wuosmaa, Preprint (1999); see also 
        B. B. Back, D. J. Blumenthal, C. N. Davids, D. J. Henderson, 
	R. Hermann, D. J. Hofman, C. L. Jiang, H. T. Penttil\"{a}, and
	A. H. Wuosmaa,
	''Fission and Properties
	of Neutron-rich Nuclei'', World Scientific, 1998, 
	eds. Hamilton and Ramayya, p. 192.
        
\bibitem{Hofman95}
        D. J. Hofman, B. B. Back, and P. Paul,
	Phys. Rev. {\bf C51} (1995) 2597.
	
\bibitem{Rubehn}
        Th. Rubehn, K. X. Jing, L. G. Moretto, L. Phair, K. Tso, and 
	G. J. Wozniak,
	Phys. Rev. {\bf C54} (1996) 3062.
\end{thebibliography}
\end{document}